# Reinventing Solid State Electronics: Harnessing Quantum Confinement in Bismuth Thin Films

Farzan Gity[1], Lida Ansari[1], Martin Lanius[2], Peter Schüffelgen[2], Gregor Mussler[2], Detlev Grützmacher[2], J. C. Greer[1]*

[1] Tyndall National Institute, Lee Maltings, Dyke Parade, Cork, Ireland T12 R5CP

[2] Peter Grünberg Institute 9 & Jülich Aachen Research Alliance (JARA-FIT), Research Center Jülich, Germany

*Jim.Greer@tyndall.ie

Solid state electronics relies on the intentional introduction of impurity atoms or dopants into a semiconductor crystal and/or the formation of junctions between different materials (heterojunctions) to create rectifiers, potential barriers, and conducting pathways. With these building blocks, switching and amplification of electrical currents and voltages is achieved. As miniaturization continues to ultra-scaled transistors with critical dimensions on the order of ten atomic lengths, the concept of doping to form rectifying junctions fails and heterojunction formation becomes extremely difficult. Here it is shown there is no need to introduce dopant atoms nor is the formation of a heterojunction required to achieve the fundamental electronic function of current rectification. Ideal diode behavior or rectification is achieved for the first time solely by manipulation of quantum confinement in approximately 2 nanometer thick films consisting of a single atomic element, the semimetal bismuth. Crucially for nanoelectronics, this new quantum approach enables room temperature operation.

**One Sentence Summary:** Quantum confinement in a semimetal bismuth film is exploited to perform current rectification near atomic scale limits and with room temperature operation.

Dennard or linear scaling of transistors to reduce costs while enhancing performance (*1*), now known as the 'happy scaling' era, came to an end during the mid-1990s for integrated circuit manufacture. Since that time, various technology boosters have been implemented to overcome what were incorrectly foreseen as insurmountable barriers. Technology boosters include the use of new material combinations to reduce quantum mechanical tunneling in a transistor's 'OFF' state, the intentional introduction of strain to enhance electron and hole mobilities to increase currents in the 'ON' state, and the use of silicon-on-insulator and/or three dimensional finFET structures to reduce short channel effects (*2*). Enormous scientific effort and financial investments have been made to overcome the fundamental limitations associated with the physical behaviour of materials and transistor architectures to enable continued scaling. These efforts have allowed the steady cadence for electronics miniaturization first observed during the 1960s and as expressed by Moore's 'law' (*3*). These advances have led to transistor critical dimensions of less than 10 nm, or as expressed in number of atoms, approximately 20 silicon atoms or less. At such small scales, fundamental physical principles such as diffusion, solid-solubility limits, and statistical fluctuations pose challenges to manufacture at the length scales that industry seeks for increased functionality at lower power to fuel the improved performance that consumers and industries have come to expect with a concomitant lowering of cost.

For 'end-of-the-roadmap' transistors, the number of atoms can be on the order of a few hundred to thousand and the introduction of even a few dopant atoms introduces extremely high doping levels. Ultra-sharp doping concentration gradients to form *p-n* junctions as required in conventional transistor designs is difficult to achieve at such dimensions, and random effects due to the positions of dopant atoms suggest that such *p-n* junctions cannot be routinely realized on the scale of few nanometers. The very concept of a *p-n* junction breaks down (*4*). An alternative path forward is to eliminate the need for doping nanostructures by exploiting quantum effects that only arise on the nanoscale (*5*). Semimetals may be thought of as semiconductors with a either a 'zero or negative' forbidden energy gap or band gap. As a result, quantum confinement in low-dimensional (2D, 1D) semimetal materials leads to a semimetal-to-semiconductor transition as the physical size of a nanostructure becomes comparable to the Fermi wavelengths of electrons and holes (*6*). Bulk bismuth has a rhombohedral crystal structure, which can be expressed in terms of a hexagonal unit cell, and is a semimetal with band overlap between the valence and conduction band of 38 meV and 98 meV at 2 K and 300 K, respectively (*7-10*). Here, the quantum confinement effect is used to demonstrate that for (111) Bi thin films of thickness less than 6 nm, a 'positive' band gap > 100 meV emerges allowing for the formation of a metal-semiconductor junction between a thick (semimetallic) and thin (semiconducting) region in a single film. A metal-semiconductor junction can be either Ohmic, behaving as a simple resistor or it can act as current rectifying diode; in the latter case a Schottky barrier is said to have formed. The formation of a Schottky barrier with near ideal diode operation at room temperature is reported here for a *single* elemental film, namely bismuth, engineered to have two regions of differing thicknesses.

A quantum confinement induced band gap in Bi nanowires was predicted based on the electronic structure of bulk Bi and including the effects of confinement (*6*). For trigonal oriented square nanowires, a zero band gap material at approximately $52 \times 52$ nm$^2$ cross section is expect to be seen with a sharply increasing band gap for decreasing nanowire cross section. In ref. (*11*), bismuth nanowires were patterned with cross sections of $40 \times 30$ and $40 \times 50$ nm$^2$. Electrical measurements indicate that the wires were indeed below the semimetal-to-semiconductor critical dimension. Electrical conductivity changes of a few percent relative to the conductivity of the nanowires at T=270 K were consistent with a band gap of the order of 10 meV. Similar results have been shown for bismuth nanowires and thin films with critical dimensions of the order of 20 nm and larger. However, these reported band gaps are only of the order of a few tens of meV (*7,12*) which is not sufficiently large with respect to the thermal energy at room temperature for use in nanoelectronics applications.

Single crystalline bismuth nanowires were synthesized by electrochemical deposition in an aluminium oxide template in ref. (*13*). Thicker regions of diameter of 70 nm were grown in contact with narrower regions of approximately 30 nm diameter. An asymmetry in the currents by a factor of ~1.5 is found at forward and reverse bias voltages of ±1 V, which the authors propose may be analogous to a metal-degenerate semiconductor junction. Although suggestive, the small asymmetry between forward and reverse bias is too small to be of use in nanoelectronic devices and the IV characteristic cannot be described even as a first approximation by the diode equation. Here we demonstrate for much smaller critical dimensions that near ideal diode behavior can be accomplished at room temperature with a rectification ratio of ~75 at voltages of only ±0.15 V when correcting for parasitic series resistance.

To demonstrate room temperature diode behavior, thin Bi films were grown by molecular-beam epitaxy (MBE) on Si(111) wafers. Prior to the deposition, the Si substrates were chemically cleaned by the HF-last RCA procedure to remove the native oxide and passivate the surface with hydrogen. The substrates were subsequently heated in-situ to 700°C for 20 min to desorb the hydrogen atoms from the surface. The Bi material flux was generated by an effusion cell operated at a temperature of 550°C, which yields a growth rate of 17 nm/h. Thickness and crystalogrophic orientation are the key parameters determining the electronic structure of thin films. X-ray reflectivity (XRR) measurement of three exemplary Bi layers with thicknesses 28 nm, 5.7 nm, and 3.2 nm can be seen in Fig. S3(A) in the SM and Fig. S3(B) shows x-ray diffraction (XRD) curves for the three films. The XRR curves display a high number of thickness oscillations evidencing smooth surfaces and interfaces with a simulated RMS value < 0.5 nm. X-ray diffraction (XRD) for the same three layers reveal signatures at $2\theta = 22.49°$ and $45.92°$, which refer to the (111) and (222) reflections of Bi and hence indicate single crystal growth with the (111) orientation in the growth direction.

Hall measurements were performed to determine thin charge carrier mobility, carrier charge, free carrier concentrations, as well as resistivity. Temperature dependence of the carrier concentration is shown in Figure 1 for the three Bi film thicknesses of 28 nm, 5.7 nm and 3.2 nm. For the thicker Bi film, the carrier concentration is largely independent of temperature for the range shown indicating semimetallic behavior. However for the 5.7 nm film, more than an order of magnitude increase in the carrier concentration is seen over the temperature range from 200 K to 300 K revealing characteristic semiconducting behavior consistent with the onset of a quantum confinement induced band gap. Although the 3.2 nm Bi film is not single crystal (note the higher carrier concentration relative to the 5.7 nm film), it possesses a high mobility similar to the crystalline 5.7 nm film. The Hall mobility for both of the thinner films remains remarkably high as shown in Figure 1(A). Estimated free carrier concentrations obtained from DFT calculations for bulk Bi and a thin film with an induced band gap of 120 meV are also shown in Figure 1(A) with reasonable agreement to the experimental data. The temperature dependence of the resistivity for Bi thin films with varying thicknesses is shown in Figure 1(B). Considering conventional resistivity classification of metals, semimetals and semiconductors, the resistivity of the crystalline ultra-thin 5.7 nm Bi film is well within the semiconducting range and displays increased resistivity with lowering temperature.

The effects of quantum confinement and band folding determine the value of the confinement induced band gap in the (111)-oriented thin films and electronic structure calculations were performed to correlate the theoretical predictions to the band gaps extracted from the electrical measurements. Standard theoretical methods for calculating band structures based on density functional theory (DFT) are well known to suffer deficiencies with respect to predicting band gaps primarily due to approximations for the electronic exchange and correlation energies. These approximations typically result in a large and systematic underestimation of band gaps in semiconductors. The semimetallic band structure of bulk Bi found using DFT predicts a much larger band overlap of ~160 meV (*16,17*) compared to the experimental value. To provide an improved description of the confinement effect in bismuth thin films, particularly with respect to determining band gap energies, the *GW* (*G*: Green's function, *W*: screened Coulomb interaction) method in conjunction with a many body perturbation theory (MBPT) correction is used (*18*). The

*GW* electron self-energy is used to correct the single particle energies resulting from the Kohn-Sham equations used to find the electron charge density within DFT. Due to the computational demands of *GW* calculations, this approach was applied to films between approximately 1 nm and 2 nm thickness and used to extrapolate the effect of quantum confinement for thicker films. The role of surface chemistry on the band gap is investigated by considering –H and –OH terminated surfaces. In the DFT calculations without surface terminating species, the band gap disappears in films thicker than 2 bilayers due to the presence of surface states. However removing surface states by hydroxyl and hydrogen terminations, the density of states (DoS) vanishes at the Fermi level in thin films with a band gap opening for thicknesses of several nanometer. As can be seen in Figure 2, hydroxyl termination provides a slightly larger value of the band gap in comparison with hydrogen termination, an effect also found for silicon nanowires (*19,20*). The experimental samples have a native oxide layer on the Bi surfaces, therefore, the –OH termination is assumed to better describe the surface chemistry and surface charge exchange in the films. The band gap extracted from the electrical measurements of Bi film with 5.7 nm indicates a band gap of 125 meV, which is in good agreement with the theoretical predicted value of 117 meV as extracted from the thermal activation of free carriers with the assumption the semiconducting Bi region is intrinsic. Hence the electrical measurements are consistent with the theoretically predicted band gaps for (111)-oriented bismuth films with thickness < 6 nm.

To form a monomaterial rectifier using the bismuth films, an etch recipe was developed to selectively thin regions of the Bi film in a controlled manner. Ti/Au was deposited on a 12 nm semimetallic Bi film and patterned to form circular transfer length method (cTLM) electrical contacts. The metal contact structure was then used as hard mask with the native oxide removed and the Bi film thinned using an argon plasma etch. Thinning of the exposed bismuth was followed by evaporation of a 15 nm $HfO_2$ layer. This process was followed by an etch step to open the contact regions to allow for electrical measurement. Electrical measurements of the cTLM contacts on the 12 nm un-etched sample (reference sample) showed Ohmic current-voltage (IV) characteristics over a temperature range of −40 °C to 20 °C as shown in the SM yielding a contact resistance of $9\times10^{-4}$ Ω·cm$^2$. The sheet resistance obtained from the cTLM analysis of $\rho_{cTLM} \approx 8\times10^{-4}$ Ω·cm is in good agreement with the Hall measurements $\rho_{cTLM} \approx 6\times10^{-4}$ Ω·cm. To vary the thickness of the bismuth film exposed to the etch step, the etch duration was varied between 60 seconds and 80 seconds. Electrical characteristics of the thick/thin devices are shown in Figure 3(A) for the resulting 8.2 nm, 4 nm, and 1.5 nm Bi films exposed to the etch step with thickness determined by TEM. The IV characteristic of the 12 nm/8.2 nm junction is approximately temperature independent consistent with the quantum confinement effect not playing an appreciable role at 8.2 nm thick films. The dependence on temperature increases for the 12 nm/4 nm junction since the band offset at the thick/thin interface (*i.e.,* 132 meV) is not large enough to prevent thermionic emission dominating the reverse current at room temperature. As the exposed bismuth region is further thinned to approximately 1.5 to 2 nm, the formation of a Schottky barrier diode is observed through the rectifying nature of the IV characteristic. The semimetal-semiconductor barrier is further increased to ~0.34 eV for the 1.5nm film as determined from the temperature dependence of the reverse saturation current. For a midgap Fermi level alignment, the estimated band gap is ~0.70 eV which is again in good agreement with the theoretical calculations as compared to in Figure 2. This is the first time near ideal diode behavior has been observed for a diode fabricated as a monomaterial junction. The current rectification is achieved at room temperature and with the observed asymmetry in the IV characteristic comparable to macroscale diode junctions.

This study presents a key step forward toward a fundamental shift in nanoelectronics by exploiting quantum effects to mimic the operation of macroscopic microelectronics. It is revolutionary in the sense it relies on a new material set, yet this approach is capable of slotting into the highly developed machinery of integrated circuit manufacturing. The results prove that it is unnecessary to introduce dopant atoms to create a *pn* junction nor is it required to form a heterojunction to achieve the fundamental function of current rectification. This work demonstrates that the fundamental laws of quantum mechanics can be harnessed to enable new nanoelectronics innovation as electronic devices are manufactured on the length scale of tens of atoms, and crucially, this can be achieved for room temperature device operation.

**Acknowledgments**

This work has been funded by Science Foundation Ireland through the Principal Investigator award 13/IA/1956 with support from Intel Corporation through the Intel Tyndall Collaboration Program. Calculations were performed on the high performance computing clusters provided at Tyndall and the Irish Centre for High End Computation (ICHEC).We thank Dr. Michael Schmidt at Tyndall for transmission microscopy imaging.

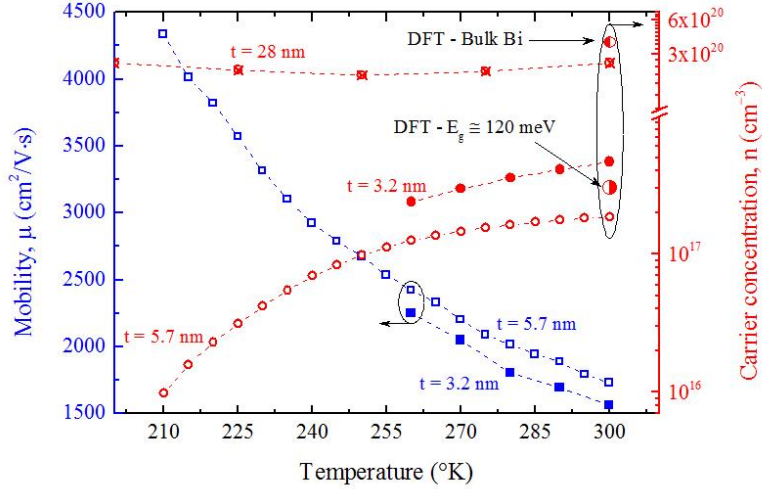

(A)

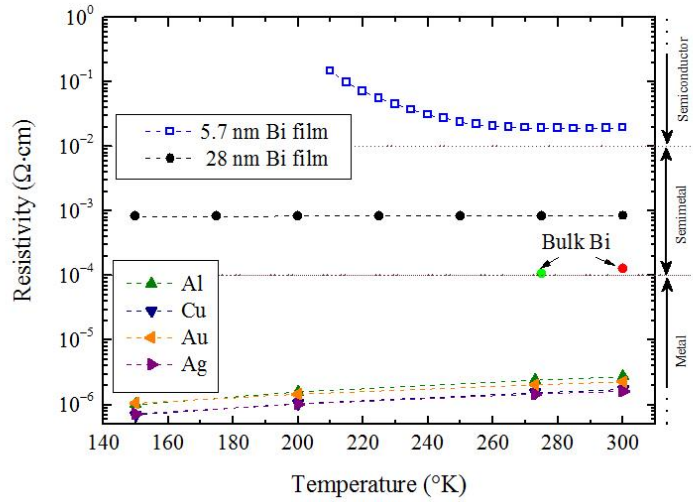

(B)

**Fig. 1**. Electrical characterization data for thin bismuth films of thickness 28 nm, 5.7 nm and 3.2 nm. The thicker films are crystalline with (111) surface orientation, the 3.2 nm film is polycrystyalline. (**A**) Temperature-dependent mobility of the 5.7 and 3.2 nm Bi [111] films (blue). Carrier concentration as a function of temperature for Bi films with thickness 28, 5.7 and 3.2 nm reveals limited variation with temperature for the thicker 28 nm film but considerable reduction with temperature for the thinner 5.7 and 3.2 nm films. Carrier concentration estimates based on DFT band structures for bulk Bi and a thin film with a band gap ≈ 120 meV are shown for comparison. Due to larger quantum confinement effect in the the 3.2 nm Bi film, the metal contact/Bi interface becomes rectifying at low temperatures hampering the electrical extraction for temperatures less than 260 K. The same electrical limitation is observed for the 5.7 nm Bi film for temperatures less than 210 K. (**B**) Resistivity of Bi films versus temperature showing for the 5.7 nm and 28 nm Bi films, respectively. Resistivity for bulk Bi at two temperatures as taken from literature (*14,15*) and for several common metals (*15*) are shown for reference.

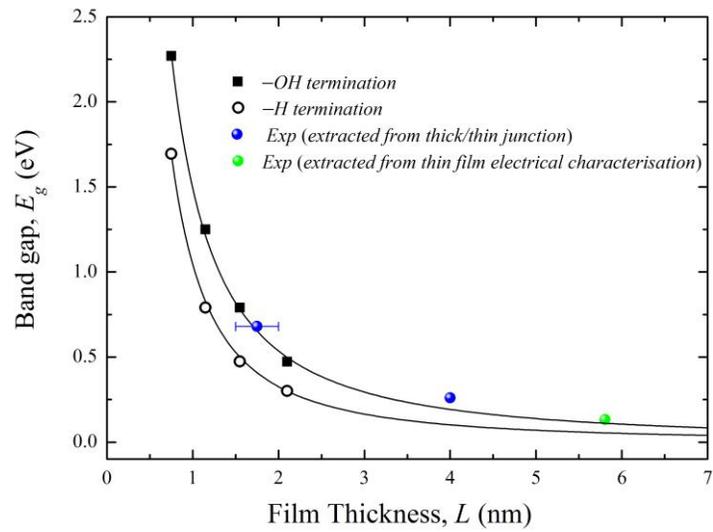

**Fig. 2**. Band gap versus film thickness and surface termination from *GW* calculations for Bi (111) slabs. The solid curves show a power law fit $\alpha/L^\beta$ where $L$ is the film thickness and $\alpha$, $\beta$ are fitting parameters. The green sphere indicates experimental data ($E_g$ = 125 meV). Blue spheres are experimental estimates based on extraction from the current-voltage characteristics at different temperatures for the thick/thin junctions.

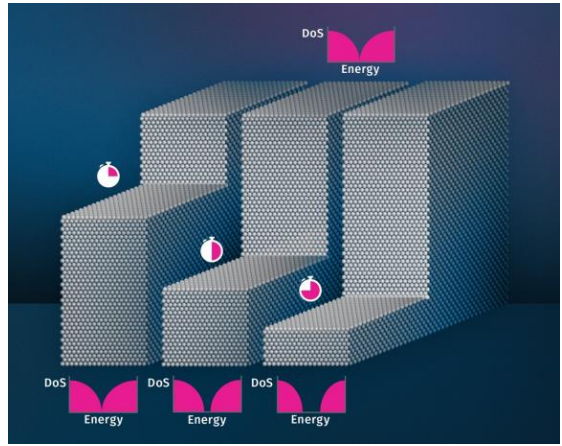

**Fig. 3.** Illustration of the "thick/thin" junctions formed with different etching times used to create a thin region. The stopwatch icon represents the proportional time for the thin region etch step. As the etched region is 'thinned' a forbidden gap between the conduction and valence bands as shown in the schematic representation of the density of states (DoS) corresponding to the fil thickness at the top and bottom of the figure emerges; detailed DoS plots for the film thicknesses considered are shown in the supplementary material.

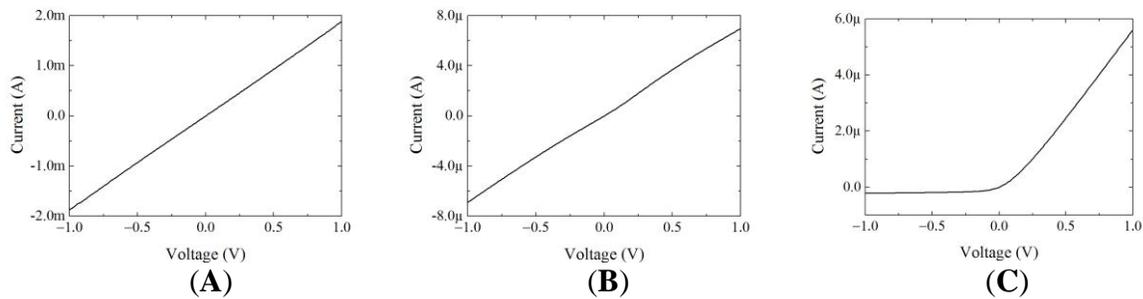

**Fig. 4.** Experimental current-voltage (IV) characteristics at T=20 °C for junction with thicknesses (**A**) 8 nm, (**B**) 4 nm, and (**C**) 1.5 nm in the thin region. Ohmic or near Ohmic behaviour is observed for the two thicker films. For the thick/thin junction in (**C**), diode behavior at room temperature is seen. As shown in the supplementary material, the IV characteristic when corrected for series resistance is consistent with exponential behavior in the forward direction. The 12nm/4nm structure (**B**) becomes rectifying at T= -40 °C as shown in the SM consistent with smaller voltage barrier arising from the smaller band gap in the thin region with respect to the structure in (**C**).

# Supplementary Materials for

## Reinventing Solid State Electronics:
## Harnessing Quantum Confinement in Bismuth Thin Films

Farzan Gity[1], Lida Ansari[1], Martin Lanius[2], Peter Schüffelgen[2], Gregor Mussler[2], Detlev Grützmacher[2], James C. Greer[1]*

[1] Tyndall National Institute, Lee Maltings, Dyke Parade, Cork, Ireland T12 R5CP

[2] Peter Grünberg Institute 9 & Jülich Aachen Research Alliance (JARA-FIT), Research Center Jülich, Germany

*Jim.Greer@tyndall.ie

**This PDF file includes:**

    Materials and Methods
    Supplementary Text
    Figs. S1 to S5
    References (*21-23*)

## Materials and Methods

Bismuth thin films were grown by molecular beam epitaxy (MBE) on Si(111) wafers. X-ray reflectivity (XRR) measurements for three exemplary Bi layers are displayed in Figure S1(A) and Figure S1(B) shows the results from X-ray diffraction (XRD) measurements for the same set of thin films. The high-resolution cross-sectional transmission electron microscopy (HR-XTEM) image of the 5.7 nm film is shown in Figure S1(C) and is consistent with the single crystal nature of the bismuth thin films.

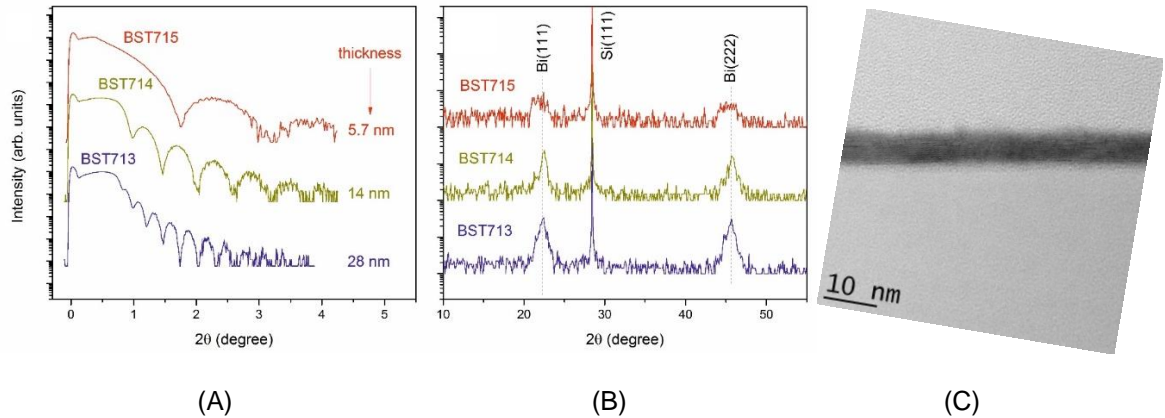

(A) (B) (C)

Fig. S1 - (A) XRR of bismuth film with thicknesses of 5.7 nm, 14 nm and 28 nm on Si (111). (B) XRD of 5.7 nm, 14 nm and 28 nm bismuth film thicknesses on Si (111). (C) HR-XTEM of 5.7 nm Bi film grown by MBE on Si(111).

## Supplementary Text

Computational details - The norm-conserving pseudopotentials utilizing the Perdew-Zunger form of local density approximation (LDA) are used in the density functional theory (DFT) calculations (*21*). For the numerical atomic orbital basis set (*22*), three orbitals per atomic level are defined with the cutoff distance of 7 Bohr. The atomistic structure is relaxed with respect to atom positions and the simulation cell parameters to minimize the total energy such that the maximum force component per atom is less than 0.01 eV/Å. It is well known that commonly applied approximations such as LDA to the exchange correlation functional in DFT underestimate band gaps in semiconductors and insulators. Electronic structure using DFT can allow for a characterization of a wide variety of materials as conductors, semiconductors, or metals. However, to provide a quantitative description of the opening of a band gap in bismuth nanowires due to quantum confinement requires a more accurate treatment of the band structure. The *GW* (*G*: Green's function, *W*: screened Coulomb interaction) approximation is a method for determining quasi-particle excitations in many-electron systems and provides an accurate prediction of band gaps that are in good agreement with experimental values (*23*). In the main text, the effects of band gap widening due to quantum confinement are described using the *GW* approximation. In the following, the DFT calculations are detailed to provide information on the simulation cells used for the calculations and to provide information on the geometry of atomistic models for thin films, details of their density of states (DoS), and the effect of surface termination on thin bismuth films.

The MBE Bi films' orientation in the growth direction is trigonal. To investigate the influence of quantum confinement on the emergence of a band gap in bismuth thin films of varying thicknesses with trigonal orientation normal to the surface, electronic structure calculations for bismuth slab models of approximately 2 nm Bi thickness are performed. Figures S2(A) and S2(B) show side views of slabs with hydroxyl (–OH) surface termination and without surface terminations for the relaxed structures. Top views of the slabs are shown in Figure S2(C). The bilayers characteristic of bulk bismuth in the [111] orientation form covalently bonded layers normal to the surface. The atoms' next-nearest neighbors are in adjacent bilayers and bonding within a bilayer is much stronger than the inter-bilayer bonding. The slab model without surface termination shows a ~3% reduction in the next-nearest neighbors' bonds lengths compared to the bulk Bi structure. The reduction for the slab model with –OH terminating group increases by ~6% while the nearest neighbor bond length is not modified significantly either with or without surface hydroxyl groups. The charge difference distribution in lateral plane is shown for the Bi thin film with –OH and without termination in Figures S2(D) and S2(E), respectively. The blue coloring indicates charge accumulation and red coloring indicates charge depletion relative to neutral atomic charge densities providing an indication of the charge reorganization occurring in the slabs for bond formation. Due to the electronegativity of –OH, there is a considerable charge transfer to the surface for the passivated slab model. The impact of stronger surface charge transfer on the formation of a band gap in the –OH passivated slab is shown in Figure S2(E) in which the DoS of the –OH terminated slab is compared with the DoS for the slab without termination and hence in the presence of surface dangling bonds. The projected DoS (PDoS) for Bi atoms located at the surface and core of the Bi slab model with and without surface passivation is plotted in Figure S2(F). As illustrated, the PDoS of atoms near the surface of the unpassivated slab varies significantly in comparison to core atoms for energies about the Fermi level. Within the –OH terminated slab model, there is not a significant difference in the PDoS between the surface and core metal atoms revealing the elimination of the surface dangling bonds.

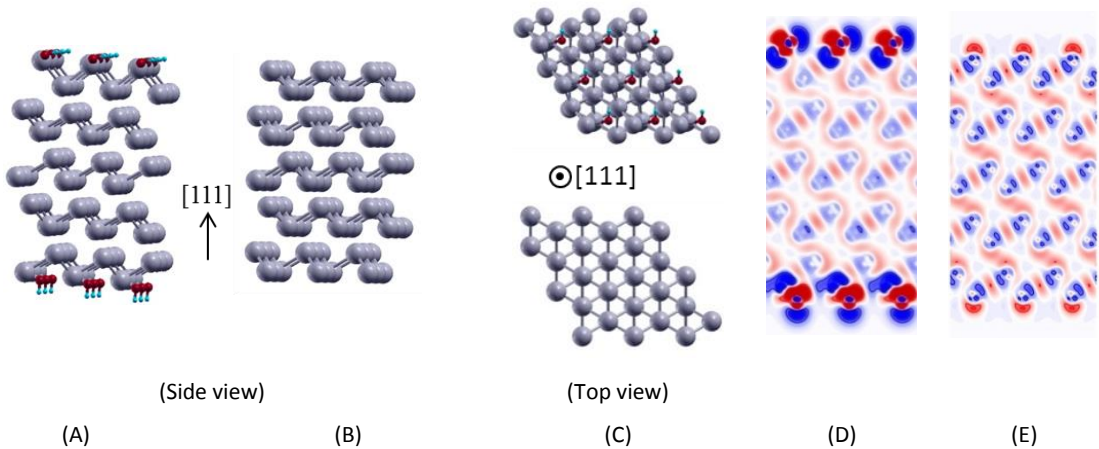

(Side view)  (Top view)

(A)  (B)  (C)  (D)  (E)

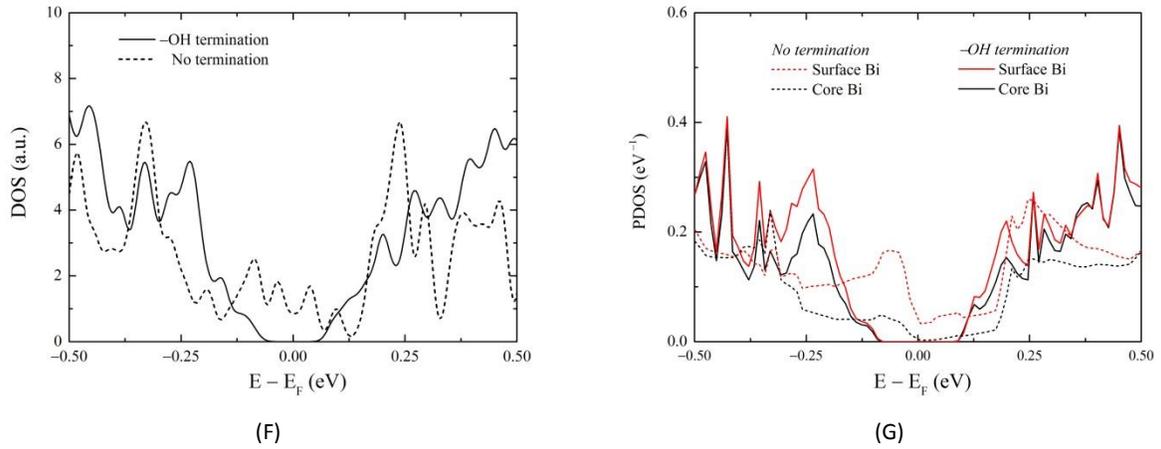

(F)   (G)

**Fig. S2** - Atomic illustration of relaxed (strain free) Bi slab models. Side view (A) with hydroxyl surface termination, and (B) without surface termination. Top view with hydroxyl (C-top) and without (C-bottom). Contour plot of lateral plane charge difference density distributions for Bi slabs for (D) hydroxyl surface termination, and (E) without. (F) Total DoS for the Bi thin films with –OH termination and without. (G) PDoS for surface and core Bi atoms with and without –OH termination.

The DoS of the etched 8nm, 4nm, and 1.5nm films (*i.e.,*) is compared to the as-grown 12 nm film illustrated in Figure S3. The band gap in thin region emerges and increases as the thickness of the film decreases.

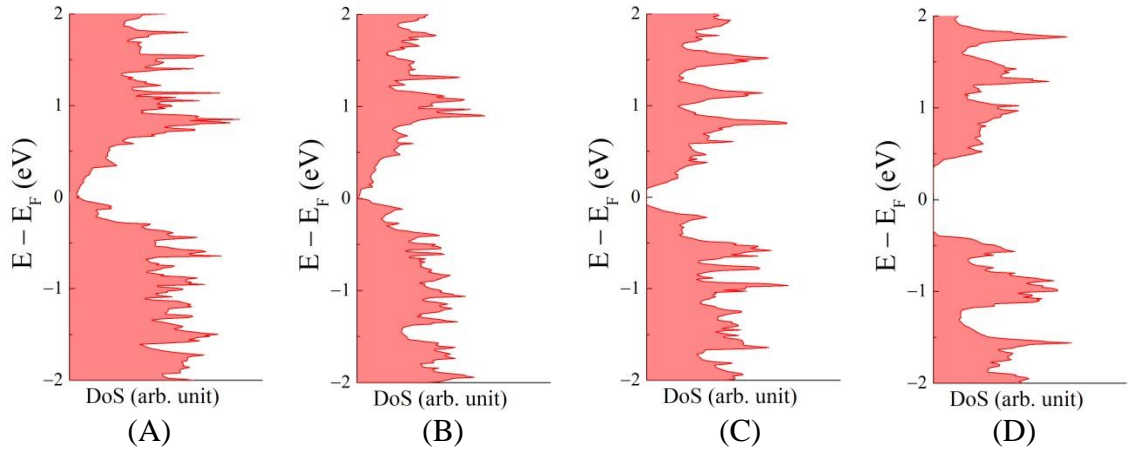

(A)   (B)   (C)   (D)

Fig. S3 – The density of states for bismuth bulk and thin films from DFT calculations corresponding to schematic band structures shown in Fig. 3 of the main text for simulation slabs of thickness (A) bulk, (B) 8 nm, (C) 4 nm, and (C) 1.5 nm with hydroxyl termination.

Electrical characterization details - Electrical measurements of the circular transmission line model (cTLM) contacts on the 12 nm un-etched sample (reference sample) show Ohmic current-voltage (IV) characteristics as presented in Figure S4.

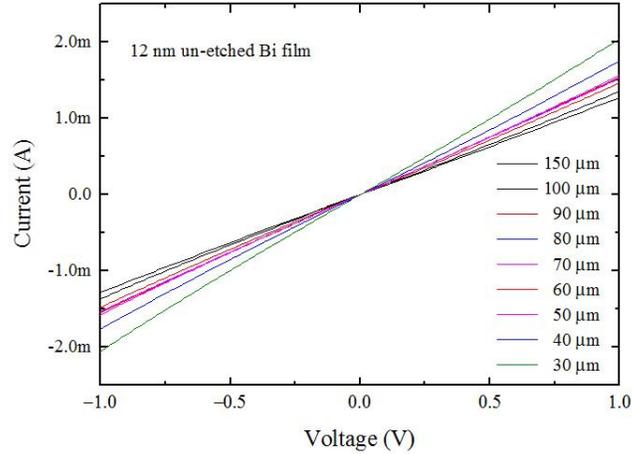

**Fig. S4** - Electrical measurements of the cTLM contacts on the 12 nm un-etched sample (reference sample) showing Ohmic current-voltage (IV) characteristics.

Current-voltage (IV) characteristic of the 12nm/1.5nm diode at room temperature is shown in Figure S5 corrected for the effect of the series resistance ($R_s$). $R_s$ and ideality factor of the diode is extracted by fitting the measured data with the diode equation. Including a series resistance of ~150 kΩ to describe the contact resistance due to the presence of Bi native oxide at the interface of the thick Bi section of the junction and metal contact. With an ideality factor of 1.82, the forward current characteristic is well described as an exponential function of applied voltage bias across the junction.

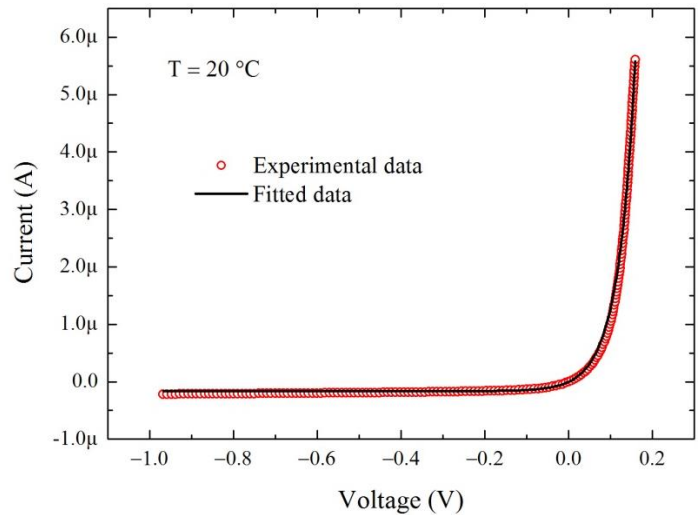

**Fig. S5** - Current-voltage (IV) characteristic of the *homo*material bismuth junction formed with film thicknesses 12nm/1.5nm operating near room temperature with the effects of parasitic series resistance of 150 kΩ removed.

Finally, in Fig. S6 the current-voltage characteristic is for the 12nm/4nm junction is shown measured at T=-40 °C. The junction becomes rectifying at lower temperatures consist with a smaller voltage barrier for the 12nm/4nm film relative to the 12nm/1.5nm junction. The 12nm/8nm film remains Ohmic for all temperatures greater than -40 °C consistent with a voltage barrier comparable to $k_BT$ at room temperature.

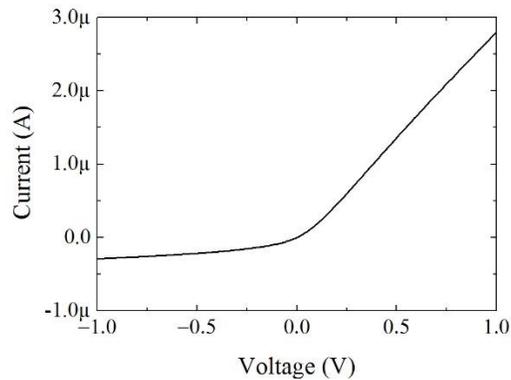

**Fig. S6** – Current voltage characteristic for the 12nm/4nm junction at T=-40 °C. The same junction is nearly Ohmic at T=+20 °C consistent with a smaller voltage barrier relative to the 12nm/1.5nm junction.